\documentclass[namedreferences]{solarphysics}
\usepackage[optionalrh]{spr-sola-addons} 
\usepackage{url}
\usepackage{subfig}
\usepackage{cite}
\usepackage[pdftex]{graphicx}
\usepackage{sidecap}
\usepackage{epstopdf}
\usepackage[pdfborder={0 0 0 },urlcolor=blue,breaklinks]{hyperref}
\usepackage{solaheader}	
\newcommand{\solareceiveddate}{10 January 2013} 
\newcommand{\solaaccepteddate}{3 June 2013}
\newcommand{\Xdoi}{10.1007/s11207-013-0344-2}
\renewcommand{\solayear}{2013}

\newcommand{\arcsec}{\hbox{$^{\prime\prime}$}}

\begin{document}
\begin{article}
\begin{opening}
\title{Effects of Stratification and Flows on P1/P2 Ratios and Anti-Node Shifts within Closed Loop Structures}
\author{R.~\surname{Erd\'elyi}$^{1}$\sep
        A.~\surname{Hague}$^{1}$\sep
C. J.~\surname{Nelson}$^{1}$$^,$$^{2}$
       }
\runningauthor{Erd\'elyi \it{et al.}}
\runningtitle{Period Ratios For Flowing Threads}
   \institute{$^{1}$ Solar Physics and Space Plasma Research Center, University of Sheffield, Hicks Building, Hounsfield Road, Sheffield, S3 7RH;\\
	      Email: (robertus; sma08abh; c.j.nelson)@shef.ac.uk\\
              $^{2}$ Armagh Observatory, College Hill, Armagh, BT61 9DG 
             }
\begin{abstract}
The solar atmosphere is a dynamic environment, constantly evolving to form a wide range of magnetically dominated structures (coronal loops, spicules, prominences, {\it etc}.)  which cover a significant percentage of the surface at any one time. Oscillations and waves in many of these structures are now widely observed and have led to the new analytic technique of solar magneto-seismology, where inferences of the background conditions of the plasma can be deduced by studying magneto-hydrodynamic (MHD) waves. Here, we generalise a novel magneto-seismological method designed to infer the density distribution of a bounded plasma structure from the relationship of its fundamental and first harmonics. Observations of the solar atmosphere have emphatically shown that stratification, leading to complex density profiles within plasma structures, is common thereby rendering this work instantly accessible to solar physics. We show, in a dynamic waveguide, how the period ratio differs from the idealised harmonic ratios prevalent in homogeneous structures. These ratios show strong agreement with recent observational work. Next, anti-node shifts are also analysed. Using typical scaling parameters for bulk flows within atmospheric waveguides, {\it e.g.}, coronal loops, it is found that significant anti-node shifts can be predicted, even to the order of $10$ Mm. It would be highly encouraged to design specific observations to confirm the predicted anti-node shifts and apply the developed theory of solar magneto-seismology to gain more accurate waveguide diagnostics of the solar atmosphere.
\end{abstract}
\keywords{Oscillations, Solar; Waves, Magnetohydrodynamic; Waves, Propagation.}
\end{opening}

\section{Introduction}
     \label{S-Introduction}
The ubiquitous magnetic fields within the solar atmosphere create a vast array of, often dynamic, plasma structures, such as coronal loops and spicules, which are known as `waveguides' for magneto-hydrodynamic (MHD) waves. Such waves have been widely studied in recent years due to the potential that they may contribute non-thermal energy to heating the corona (for a review of MHD waves, see: \opencite{Roberts00}; \opencite{Ruderman09}; \opencite{TR09}; \opencite{Mathioudakis13}). In this article we focus our attention on a thin, finite-length, magnetic string, analogous to a coronal loop, to study how density stratification effects the relationship between the fundamental and subsequent harmonics in a dynamic waveguide.

Oscillations in magnetic, cylindrical structures were extensively studied, and summarised in a recently popular form, in a theoretical capacity, by \inlinecite{Edwin83} and \inlinecite{Roberts84}. However, it was not until \inlinecite{Aschwanden99} interpreted coronal loop oscillations, viewed using the {\it Transition Region and Coronal Explorer} (TRACE) satellite, as manifestations of the fast kink-mode that they were first observed. Recent improvements in spatial and temporal resolutions have allowed a large number of oscillations to be observed, {\it e.g.}, Alfv\'{e}n waves first observed by \inlinecite{Jess09} in the lower solar atmosphere (for a review see \opencite{Mathioudakis13}); kink waves (\opencite{Aschwanden99}; \opencite{Verwichte04}; see \opencite{Andries09}, for a review); longitudinal waves (\opencite{Deforest98}; \opencite{Berghmans99}; reviews of sausage modes include \opencite{Demoortel09} and \opencite{Wang11}). In particular, \inlinecite{Verwichte04}, claimed to have observed evidence of the first harmonic of an oscillating coronal loop; a potentially important tool for solar magneto-seismology.

A plethora of examples of the fundamental and first harmonic within an active region have been reported in recent years ({\it e.g.}, \opencite{DM07}; \opencite{Oshea07}). Analysing a variety of TRACE observations, \inlinecite{VD07} reported P1/P2 ratios of 1.81, 1.58 and 1.795 for fast kink-mode oscillations within coronal loops.  \inlinecite{Sri08} used {\it Hinode} data to calculate a P1/P2 ratio of 1.68 for sausage modes in cool post-flare chromospheric loops. One suggested reason for the deviation of these values from two (the P1/P2 ratio for a homogeneous flux tube) is density stratification, where complex interactions of gravity and magnetism lead to inhomogeneity within the tube. Inhomogeneity within tubes may also lead to {\it spatial} periodicities. \inlinecite{Jess08} analysed the same data as \inlinecite{Aschwanden99}, finding spatial periodicities over length scales of, approximately, $3.5$ Mm. Observed period ratios within sunspot umbrae have been studied by \inlinecite{Campos86} (including the P1/P2 and P1/P3 ratios) finding deviations from homogeneous values, implying that analytical approximations for thin-tubes should be extended to thick-tubes. The P1/P2 ratio has been widely applied in recent years, for example, \inlinecite{Andries05} and \inlinecite{Verth08b} were able to estimate the scale height of a coronal loop; the latter even considering the significance of magnetic stratification.

This article expands the current topic of magneto-seismology, a technique which allows the inference of properties of the background plasma ({\it e.g.}, density structure and temperature) by analysing MHD waves. Magneto-seismology was first suggested by both \inlinecite{Uchida70} and \inlinecite{Rosenberg70}, and has since been expanded by \inlinecite{Roberts84}, and more recently investigated in the context of solar interior-atmospheric magnetic coupling, by \inlinecite{Erdelyi06}, to any magnetised solar plasma structure. There are several examples of this method being exploited including \inlinecite{Nakariakov01} and \inlinecite{Erdelyi08}. \inlinecite{Verth07} used spatial magneto-seismology to calculate the anti-node shift of a magnetic flux tube with a non-homogeneous density stratification. Further examples of magneto-seismology can be found in, for example, \inlinecite{Soler11} and \inlinecite{Soler12}. However, we expand upon these works by finding analytical approximations for specific density profiles that may model, and give insight, to the complex stratification of atmospheric waveguides.

\inlinecite{Diaz10} discussed a step-function density within a loop structure, modelling the large density gradient between its footpoint and apex, finding that for large density jumps a bead which is close to the centre of a magnetic structure, representing dense prominence threads, will be observed to have $P1/P2<2$. \inlinecite{Soler11} expanded this work to a flowing heavy thread, where the period ratio was calculated numerically, also finding a fall from the canonical value of two. In this work, we discuss a bead, analogous to a `blob' of plasma propagating slowly along a coronal loop such that $\rho_{\mathrm{thread}}\approx\rho_{\mathrm{bead}}$ (observed by, for example, \opencite{Ofman08}). We assume that the bead is close to the end of the tube (rather than in the centre as studied in, {\it e.g.}, \opencite{Diaz10}),  as the largest divergence from the homogeneous harmonic ratio should occur in spatial positions close to the end of the loop. We begin by formulating an analytical solution for a light bead which is stationary and situated close to the end of the loop. This is then expanded so that the bead may move slowly away from the end of the loop, simulating flows observed within coronal loop structures.  A comparison between these results and numerical results is also made. 

Recent high-cadence, high-resolution observations have led to discussions relating to the idea of flows within coronal loops (see, \opencite{Kopp85}; \opencite{Ofman08}). Often, localised brightenings can be observed progressing along loop structures from the footpoints in the photosphere and chromosphere into the corona (see, {\it e.g.}, \opencite{Ofman08}). How these localised flows influence MHD modes within loop structures has initially been studied in, {\it e.g.}, \inlinecite{Soler11} and \inlinecite{Soler12} in ideal MHD and by \inlinecite{Morton09} for cooling coronal loops. It has been found that small, propagating density structures within loops can cause a deviation from the expected canonical $P1/P2$ ratio. We study such a system. Further, the topic of anti-node shifts is also discussed, finding that non-homogeneous density profiles lead to large divergence from expected anti-node positions. The underpinning idea here is that eigenfunctions may be more sensitive to small ({\it i.e.}, linear) perturbations when compared to variations in eigenvalues ({\it i.e.}, frequencies) of MHD waveguides.

Our work will be structured in the following way. In Section 2, we describe the basic model. Section 3 discusses the effects of a single density discontinuity on the relationship of the harmonics of an oscillating thin tube. In Section 4, we derive the effects of a step-function density on the relationship between the harmonics of an oscillating tube before expanding this to a moving bead in Section 5. In Section 6, the anti-node shift of the harmonics is discussed in the context of spatio magneto-seismology. Section 7 draws together our conclusions and suggests further research.

\begin{figure}[!tbp]
\begin{center}
\rotatebox{0}{\includegraphics[scale=0.31,angle=270]{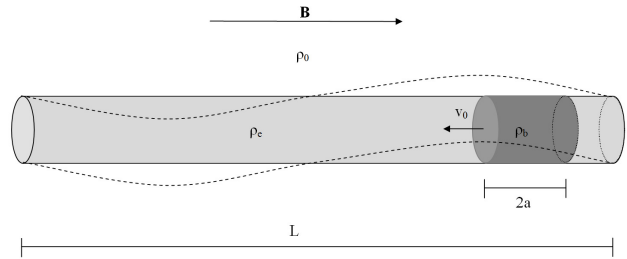}}
\end{center}

\caption{A schematic diagram of the model presented. For Section 4, $v_{\mathrm{0}}=0$. The dashed line represents the first harmonic standing oscillation of a homogeneous loop.}
\end{figure}

\section{Model Setup}
     \label{S- Model Setup}
The model which is studied can be summarised as the oscillations of a straightened, thin flux tube of length $L$ and radius $R$. We use a cylindrical coordinate system $(r,\varphi,z)$ where the $z$-axis corresponds to the axis of the tube. The magnetic field both inside and outside the tube has the form ${\mathbf{B}}=B{\mathbf{e}}_{\mathrm{z}}$ where $B$ is homogeneous with respect to $z$. As in many coronal models, the $\beta=0$ ({\it i.e.}, cold plasma) condition is applied, thereby removing any dependency on temperature. The loop has footpoints fixed in the photosphere, {\it i.e.}, its ends are immovable at $z=0$ and $z=L$.

We state that $\rho_{\mathrm{e}}$ is the background density of the tube and $\rho_{\mathrm{b}} (=\rho_{\mathrm{e}}+\delta\rho$ where $|\delta\rho|\ll |\rho_{\mathrm{e}}|)$ is the density within the bead. The tube sits within a quiet coronal plasma with density, $\rho_{\mathrm{o}}$, much less than the tube density {\it i.e.} $\rho_{\mathrm{o}} / \rho_{\mathrm{e}} \ll 1$. The bead moves with a constant speed, $v_{\mathrm{0}}$.  The setup of the problem is shown in Figure 1. This model represents an oscillating coronal loop  in which plasma, of a higher density than that of the loop, moves along the magnetic field lines, modelling observations by, {\it e.g.}, \inlinecite{Ofman08}.

The density discontinuity in our model is situated such that it is close to the end of the loop, therefore, enacting the largest influence on the oscillations of the tube (see, {\it e.g.}, \opencite{Soler11}). By specifying the placement of the bead in such a way, as well as considering a small density change, we find significantly different results from other researches in this topic ({\it e.g.} \opencite{Diaz10}; \opencite{Soler11}), including elegant analytical approximations for all harmonics. It should be noted that we consider only piecewise constant density profiles, as opposed to continuous stratification. Magnetic atmospheres with continuous stratification have been extensively studied by many authours including \inlinecite{FP58}, \inlinecite{LB79} and \inlinecite{Zhugzhda84}, among others, and has recently been extended to flux tube geometry by, {\it e.g.}, \inlinecite{Verth08a}. Geometries similar to these are thought to be more realistic of solar structures. However, a combination of flows and continuous density profiles present numerous difficulties which are not studied in the present article. The expansion of the methods presented in this research to more complex geometries could form the basis of an interesting future study.

It is assumed that the tube exists such that it can be modelled by the thin-tube approximation (TT) outlined by \inlinecite{Dymova05} {\it i.e.} $R/L\ll1$. Assuming a typical coronal loop structure, we estimate that $L\approx50-150$ Mm and $R\approx1-2$ Mm giving $R/L\approx0.007-0.04$ and, hence, that this approximation holds for our model. From this approximation, \inlinecite{Dymova05} found that the governing equation (which has later been used by, {\it e.g.}, \opencite{Erdelyi07}; \opencite{Verth08a}) of transversal perturbations can be written as
\begin{equation}
\frac{\partial^2{v(z,t)}}{\partial{t}^2}-v_{\mathrm{k}}^2(z,t)\frac{\partial^2{v(z,t)}}{\partial{z}^2}=0,
\end{equation}
where $v(z,t)$ is the transverse velocity at the tube boundary, and $v_{\mathrm{k}}(z,t)$ is the kink-speed defined as
\begin{equation}
v_{\mathrm{k,e}}(z,t)=\sqrt{\frac{2B^2}{\mu(\rho_{\mathrm{e}}+\rho_{\mathrm{o}})}}\quad\mathrm{and}\quad
v_{\mathrm{k,b}}(z,t)=\sqrt{\frac{2B^2}{\mu(\rho_{\mathrm{b}}+\rho_{\mathrm{o}})}}.
\end{equation}
Note, as $\delta\rho\rightarrow{0}$, it is simple to see that $v_{\mathrm{k,e}}(z,t)\approx{v}_{\mathrm{k,b}}(z,t)$. 

Equation (1) is only strictly applicable when $v_{\mathrm{k}}$ is independent of $t,$ {\it i.e.} there are no bulk motions present. The correct equation, taking into account bulk flows, was derived by \inlinecite{Ruderman10}; however, in a suitable limit ($v_{\mathrm{0}} \ll v_{\mathrm{A}}$, the Alfv\'en speed), Equation (1) is applicable and has been widely used by, {\it e.g.}, \inlinecite{Soler11}.

\section{Single Density Discontinuity}
     \label{S-A Single Density Discontinuity}
We begin our analysis by considering a single density jump; this can be modelled using a Heaviside function. Let us assume that the density, $\rho$, follows the distribution
\begin{equation}
\rho(z)=\left\{\begin{array}{ll}
\rho_{\mathrm{e}} & \quad{z\in[0,z_{\mathrm{0}}]};\\
\rho_{\mathrm{b}} & \quad{z\in[z_{\mathrm{0}},L]},
\end{array}\right.
\end{equation}
where $z=z_{\mathrm{0}}$ is the position of the density jump and $\rho_{\mathrm{e}}<\rho_{\mathrm{b}}$ implies $0<\delta\rho$ (or $\rho_{\mathrm{b}}<\rho_{\mathrm{e}}$ that $\delta\rho<0$).

To calculate the dispersion relationship for standing waves with these background conditions, we will use the separation of variables technique on the governing equation [Equation (1)], assuming that $v(z,t)=Z(z)T(t).$  The spatial dependence, $Z(z)$, is denoted by
\begin{equation}
Z(z)=\left\{\begin{array}{ll}
Z_{\mathrm{e}}(z) & \quad{z\in[0,z_{\mathrm{0}}]};\\
Z_{\mathrm{b}}(z) & \quad{z\in[z_{\mathrm{0}},L]},
\end{array}\right.
\end{equation}
where
\begin{equation}
Z_{\mathrm{e}}(z)=A\sin\left(\frac{\omega{z}}{v_{\mathrm{k,e}}}\right)
\end{equation}
and
\begin{equation}
Z_{\mathrm{b}}(z)=C\left[\sin\left(\frac{\omega{z}}{v_{\mathrm{k,b}}}\right)-\tan\left(\frac{\omega{L}}{v_{\mathrm{k,b}}}\right)\cos\left(\frac{\omega{z}}{v_{\mathrm{k,b}}}\right)\right],
\end{equation}
are calculated using the boundary conditions. We shall define two further conditions on the oscillating loop
\begin{equation}
\left[Z(z)\right]^{z=z_{\mathrm{0}}+\delta z}_{{z=z_{\mathrm{0}}-\delta z}}=\left[Z'(z)\right]^{{z=z_{\mathrm{0}}+\delta z}}_{{z=z_{\mathrm{0}}-\delta z}}=0, \quad\mathrm{as}\quad \delta z \rightarrow 0,
\end{equation}
thereby forcing continuity of the waveguide at $z_{\mathrm{0}}$. Applying these conditions to Equations (5) and (6) gives us the dispersion relationship,
$$
\tan\left(\frac{\omega{L}}{v_{\mathrm{k,b}}}\right)\left[\tan\left(\frac{\omega{z_{\mathrm{0}}}}{v_{\mathrm{k,e}}}\right)\tan\left(\frac{\omega{z_{\mathrm{0}}}}{v_{\mathrm{k,b}}}\right)+\left(\frac{v_{\mathrm{k,b}}}{v_{\mathrm{k,e}}}\right)\right]
$$
\begin{equation}
\quad\quad\quad\quad\quad\quad+\left[\tan\left(\frac{\omega{z_{\mathrm{0}}}}{v_{\mathrm{k,e}}}\right)-\left(\frac{v_{\mathrm{k,b}}}{v_{\mathrm{k,e}}}\right)\tan\left(\frac{\omega{z_{\mathrm{0}}}}{v_{\mathrm{k,b}}}\right)\right]=0.
\end{equation}

In order to reduce the trigonometric equation [Equation (8)] to a simpler approximation of $\omega$, dimensionless parameters, as in \inlinecite{Verth07}, are introduced. This allows us to replace the characteristic speed, the eigenfrequency, $\omega$, and $z_{\mathrm{0}}$ with their scaled counterparts
\begin{equation}
\kappa=\left(\frac{v_{\mathrm{k,b}}}{v_{\mathrm{k,e}}}\right),\quad
\gamma=\frac{\omega L}{v_{\mathrm{k,e}}},\quad
\epsilon=\frac{L-z_{\mathrm{0}}}{L}.
\end{equation}
Using the dimensionless parameters [Equation (9)] and Equation (8), we are able to calculate
\begin{equation}
\tan\left(\frac{\gamma}{\kappa}\right)=\frac{\kappa\tan\left(\frac{\omega_{\mathrm{n}}{z_{\mathrm{0}}}}{v_{\mathrm{k,b}}}\right)-\tan\left(\frac{\omega_{\mathrm{n}}{z_{\mathrm{0}}}}{v_{\mathrm{k,e}}}\right)}{\tan\left(\frac{\omega_{\mathrm{n}}{z_{\mathrm{0}}}}{v_{\mathrm{k,e}}}\right)\tan\left(\frac{\omega_{\mathrm{n}}{z_{\mathrm{0}}}}{v_{\mathrm{k,b}}}\right)+\kappa},
\end{equation}
\begin{equation}
\tan\left(\gamma\epsilon\right)=\frac{\tan\left(\gamma\right)-\tan\left(\frac{\omega_{\mathrm{n}}{z_{\mathrm{0}}}}{v_{\mathrm{k,e}}}\right)}{1+\tan\left(\gamma\right)\tan\left(\frac{\omega_{\mathrm{n}}{z_{\mathrm{0}}}}{v_{\mathrm{k,e}}}\right)},
\end{equation}
and
\begin{equation}
\tan\left(\frac{\gamma\epsilon}{\kappa}\right)=\frac{\tan\left(\frac{\gamma}{\kappa}\right)-\tan\left(\frac{\omega_{\mathrm{n}}{z_{\mathrm{0}}}}{v_{\mathrm{k,b}}}\right)}{1+\tan\left(\frac{\gamma}{\kappa}\right)\tan\left(\frac{\omega_{\mathrm{n}}{z_{\mathrm{0}}}}{v_{\mathrm{k,b}}}\right)}.
\end{equation}
After some further algebra, Equation (8) can be cast into the following, more compact, form,
\begin{equation}
\left[\tan(\gamma)-\tan(\gamma\epsilon)\right]+\kappa\left[\tan\left(\frac{\gamma\epsilon}{\kappa}\right)+\tan\left(\frac{\gamma\epsilon}{\kappa}\right)\tan(\gamma)\tan(\gamma\epsilon)\right]=0.
\end{equation}

\begin{figure}[!tbp]
\subfloat[$\epsilon$ for $\kappa=0.9$.]{\includegraphics[scale=0.30]{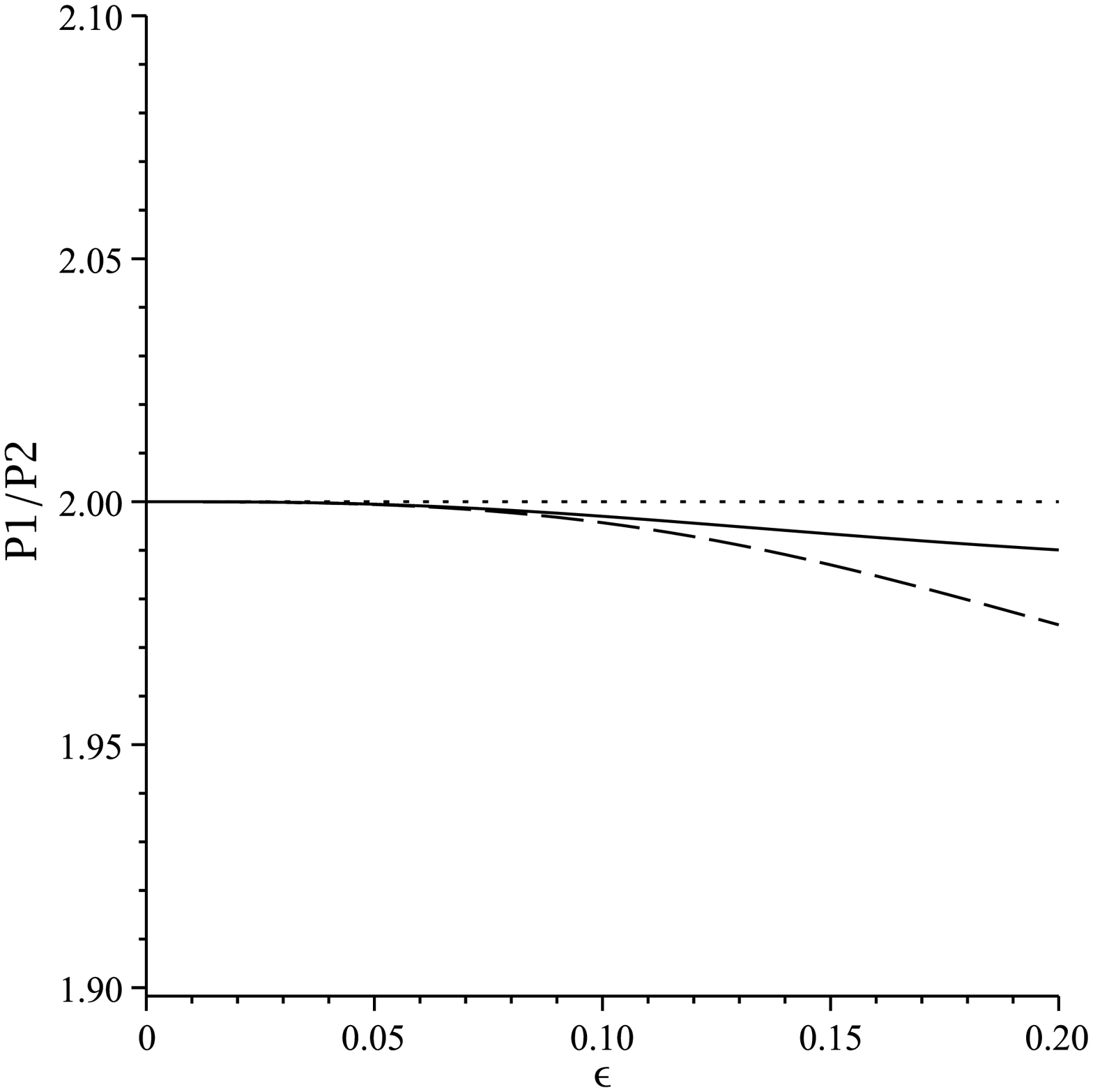}}
\subfloat[$\kappa$ for $\epsilon=0.1$]{\includegraphics[scale=0.30]{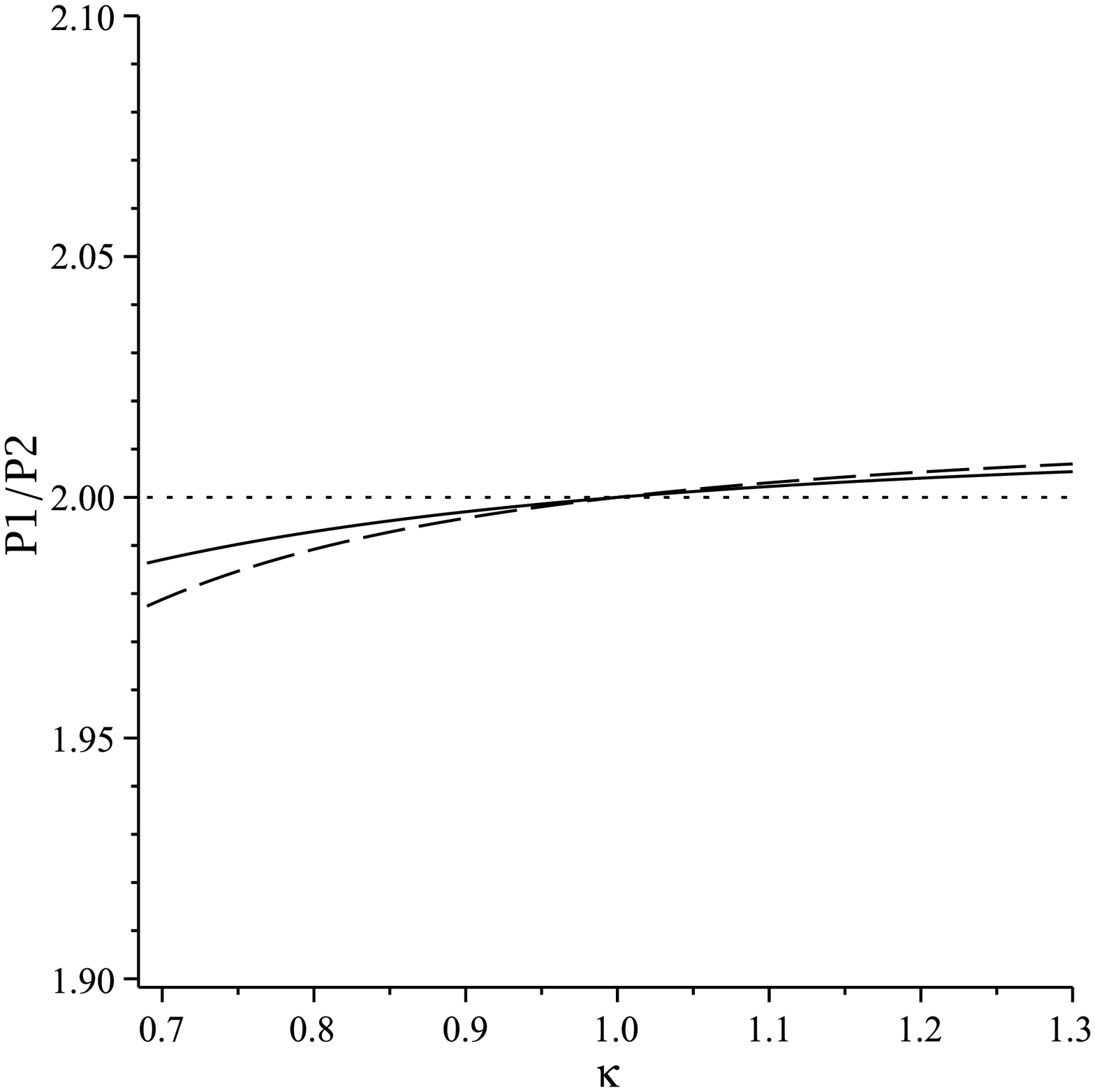}}
\caption{Fits of the analytical solution derived in Equation (20) to numerical models.}
\end{figure}

Assuming a weak stratification, such that $\rho_{\mathrm{e}}\approx\rho_{\mathrm{b}}$, it is simple to see that $\kappa\approx{1}$. Equation (10) implies
\begin{equation}
\tan(\gamma)\approx{0},
\end{equation}
and so
\begin{equation}
\gamma\approx{n}\pi, \quad n=1,2,3,...
\end{equation}
Acknowledging the basic model setup of the geometry of the problem, we limit the spatial position of the discontinuity such that $\epsilon\ll1$. Simple calculations then lead us to
\begin{equation}
\tan(\gamma)\approx\gamma-n\pi,
\end{equation}
\begin{equation}
\tan(\gamma\epsilon)\approx\gamma\epsilon+\frac{1}{3}(\gamma\epsilon)^3,
\end{equation}
\begin{equation}
\tan\left(\frac{\gamma\epsilon}{\kappa}\right)\approx\frac{\gamma\epsilon}{\kappa}+\frac{1}{3}\left(\frac{\gamma\epsilon}{\kappa}\right)^3,
\end{equation}
which allow us to rewrite Equation (13) as a cubic function in $\gamma$. Since $\epsilon$ is, by definition, small, $O(\epsilon^4)$ or higher order terms have been neglected from Equations (17) and (18). Our approximations allow the calculation of the dispersion relationship
\begin{equation}
\epsilon^2\left[\frac{1}{3}\left(\epsilon-\epsilon\kappa^2\right)+\kappa^2\right]\gamma^3-\left(\epsilon\kappa\right)^2n\pi\gamma^2+
\kappa^2\gamma-n\pi\kappa^2=0.
\end{equation}
The dispersion relation is now in a convenient form, as a wide range of methods are available for solving cubic equations, allowing Equation (19) to be solved for $\gamma$. Using the perturbation $\gamma_n=n\pi+\tilde{\gamma}$, we are able to analytically determine $\gamma$, {\it i.e.}
\begin{equation}
\gamma_n=n\pi+\frac{\epsilon^3n^3\pi^3(\kappa^2-1)}{3(n^2\pi^2\epsilon^3(1-\kappa^2)+\epsilon^2\kappa^2n^2\pi^2+\kappa^2)},
\end{equation}
hence, allowing the calculation of the $P1/P2$ ratios for this density profile (as $\gamma$ is related to $\omega$ through Equation (9), it is easy to see that $P_n \propto 1/\gamma_n$). Note that we have added the subscript $n$ to $\gamma$ to denote the $(n-1)$-th  harmonic clearly.  

We are easily able to note, that $\tilde{\gamma}$ causes the deviation of the ratios of the harmonics for $\kappa\neq1$, from the counterparts in uniform plasma. Let us plot $P1/P2$ by fixing $\kappa$ at, {\it e.g.}, $0.9$ and varying $\epsilon$ (Figure 2a)  and fixing $\epsilon$ at, say, $0.1$ while varying $\kappa$ (Figure 2b). These choices of $\kappa$ and $\epsilon$ represent a coronal loop embedded in the solar atmosphere. The period ratio is now calculated using Equation (20) (solid line); also plotted is the period ratio  computed using numerical methods with respect to Equation (13) (dashed line). A strong correlation between the numerical and analytical solutions is observed for $\epsilon <0.2$ (Figure 2a). This agrees with the assumption that $\epsilon \ll 1$. For $\epsilon=0.1$, the fit of the analytical and numerical solutions around $\kappa=1$ shows less than one percent error (Figure 2b). Note that Figure 2b shows that for $\kappa >1$, a lower density after the jump, $P1/P2>2.$ For $\kappa<1$, corresponding to a higher density after the jump, $P1/P2<2;$ as in many observed kink mode oscillations of coronal loops ({\it e.g.}, \opencite{VD07}). This model is rather basic and is not an accurate representation of observed density profiles. However,  in this work, it serves as a basis to expand the analysis in future sections.

\section{Step Function Density}
     \label{S-A Step Function Density}
Step function densities are models which can be used to approximate a range of observed phenomena within coronal loops. For example, a small bead propagating from the footpoints in the photosphere or chromosphere to the corona along a waveguide (see \opencite{Ofman08}) or stratification, where a two-layer atmosphere is chosen ({\it i.e.} a heavy photosphere and light corona leading to a decrease in density in the centre of the tube). The first example is analysed in this article. The analysis presented, focuses on a density profile defined as
\begin{equation}
\rho(z)=\left\{\begin{array}{ll}
\rho_{\mathrm{e}} & \quad{z\in[0,z_{\mathrm{0}}-a]\cup[z_{\mathrm{0}}+a,L]};\\
\rho_{\mathrm{b}} & \quad{z\in[z_{\mathrm{0}}-a,z_{\mathrm{0}}+a]},
\end{array}\right.
\end{equation}
where $z_{\mathrm{0}}$ is the center of the step function and $a$ is half the length of the step. Performing a similar analysis to Section 3, the dispersion relationship for this density profile can be written as
$$
\frac{2\gamma{\alpha}}{\kappa^2}\left[\tan^2\left(\frac{\omega{z_{\mathrm{0}}}}{v_{\mathrm{k,e}}}\right)-\tan\left(\frac{\omega{z_{\mathrm{0}}}}{v_{\mathrm{k,e}}}\right)\tan(\gamma)\right]+2\gamma{\alpha}\left[1+\tan\left(\frac{\omega{z_{\mathrm{0}}}}{v_{\mathrm{k,e}}}\right)\tan(\gamma)\right]
$$
\begin{equation}
+\left[-2\gamma{\alpha}+\tan(\gamma)\right]\left(1+\tan^2\left(\frac{\omega{z_{\mathrm{0}}}}{v_{\mathrm{k,e}}}\right)\right)=0,
\end{equation}
where
\begin{equation}
\tan\left(\frac{\omega{z_{\mathrm{0}}}}{v_{\mathrm{k,e}}}\right)\equiv \frac{\tan(\gamma)-\tan(\gamma\epsilon)}{1+\tan(\gamma)\tan(\gamma\epsilon)},
\end{equation}
and
\begin{equation}
\alpha=\frac{a}{L}.
\end{equation}
As we define the bead such that it is small, therefore $a\ll{L}$, it is simple to note that $\alpha\ll1$. This is consistent with assumptions made in other analytical works ({\it e.g.}, \opencite{Diaz10}).

Continuing the analysis as in Section 3, we are able to express $\gamma$ as
\begin{equation}
\gamma_n=n\pi+\frac{2n^3\pi^3\alpha(\kappa^2-1)\epsilon^2}{2n^2\pi^2\alpha\epsilon(\kappa^2-1)(1-3\epsilon)+\kappa^2(1+n^2\pi^2\epsilon^2)}.
\end{equation}
Figure 3 shows the period ratio using Equations (22) and (25) for the numerical and analytical solutions, respectively. As in Figure 2, we see a reasonably good agreement between the two solutions. For this plot, the newly defined length of the bead to the length of the loop ratio is held fixed at $\alpha=0.1$. Recalling the geometry of the problem (Figure 1), $\alpha=0.1$ creates the limit $\epsilon >0.1.$ Intuitively, varying $\alpha$ will give different limits for $\epsilon$. Once again we see that if $\kappa<{1}$, the ratio of the period of the harmonics drops below two (Figure 3b), {\it i.e.}, the period ratio of a homogeneous oscillating loop. This finding agrees with both intuition and with observational evidence which suggest that heavy plasma propagating along coronal loops may influence each harmonic individually.

\inlinecite{Diaz10} considered a step density function close to the centre of a loop using a comparable TT approximation. Analytical period ratios were determined for the cases for small frequency, $\gamma\ll\pi/2$ in our notation, and for a small thread $\alpha\ll1$. The main applications of the work by \inlinecite{Diaz10}, however, were towards prominence threads, principally focusing on oscillations within these structures. This, therefore, provides important differences from the model presented in this current work. For example, the present work discusses a relatively light bead positioned towards a footpoint of a loop, rather than a heavy central bead, allowing the derivation of a number of analytical solutions for periods of standing kink waves.

\begin{figure}[!tbp]
\subfloat[$\epsilon$ for $\kappa=0.9$.]{\includegraphics[scale=0.30]{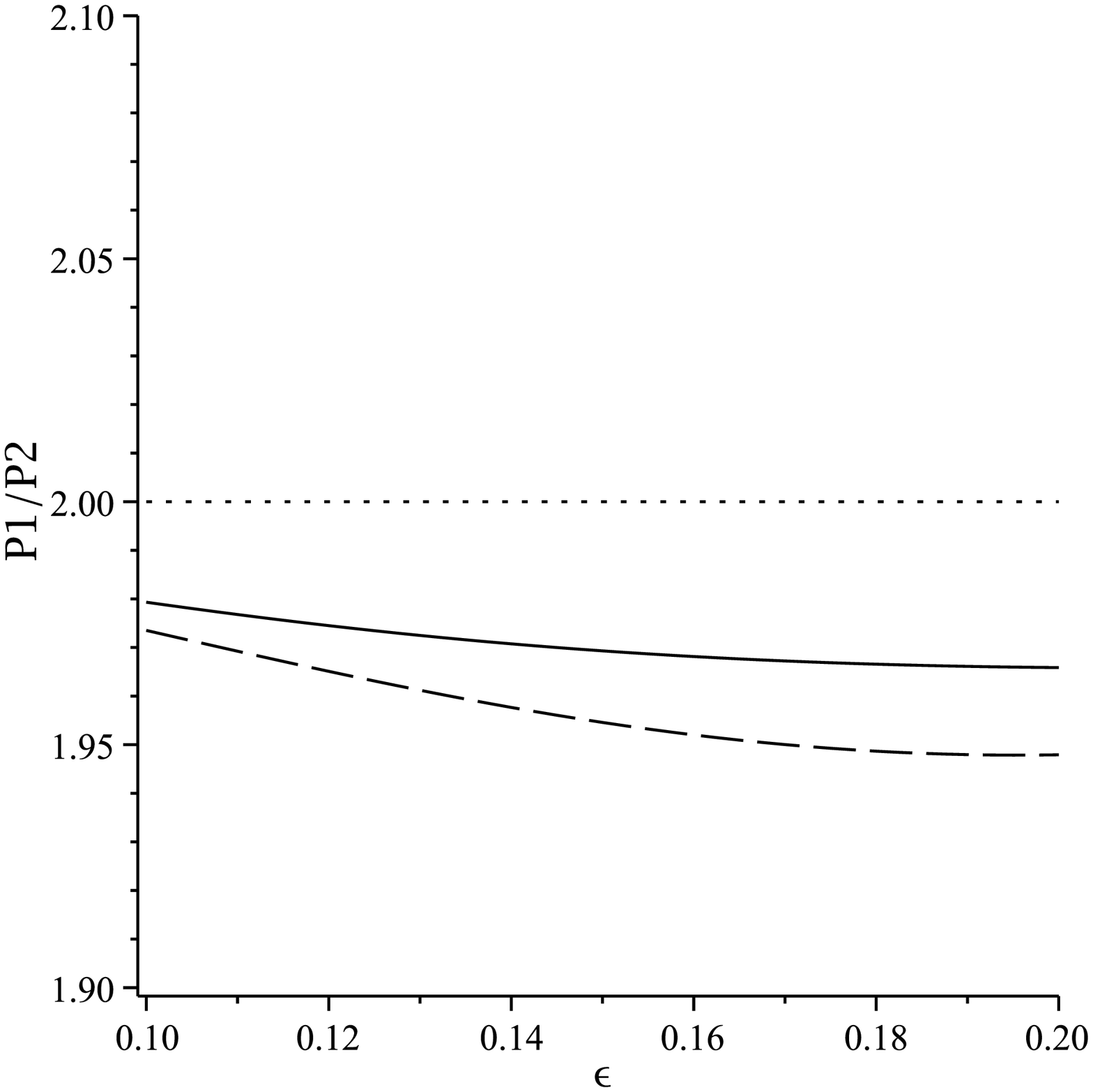}}
\subfloat[$\kappa$ for $\epsilon=0.1$]{\includegraphics[scale=0.30]{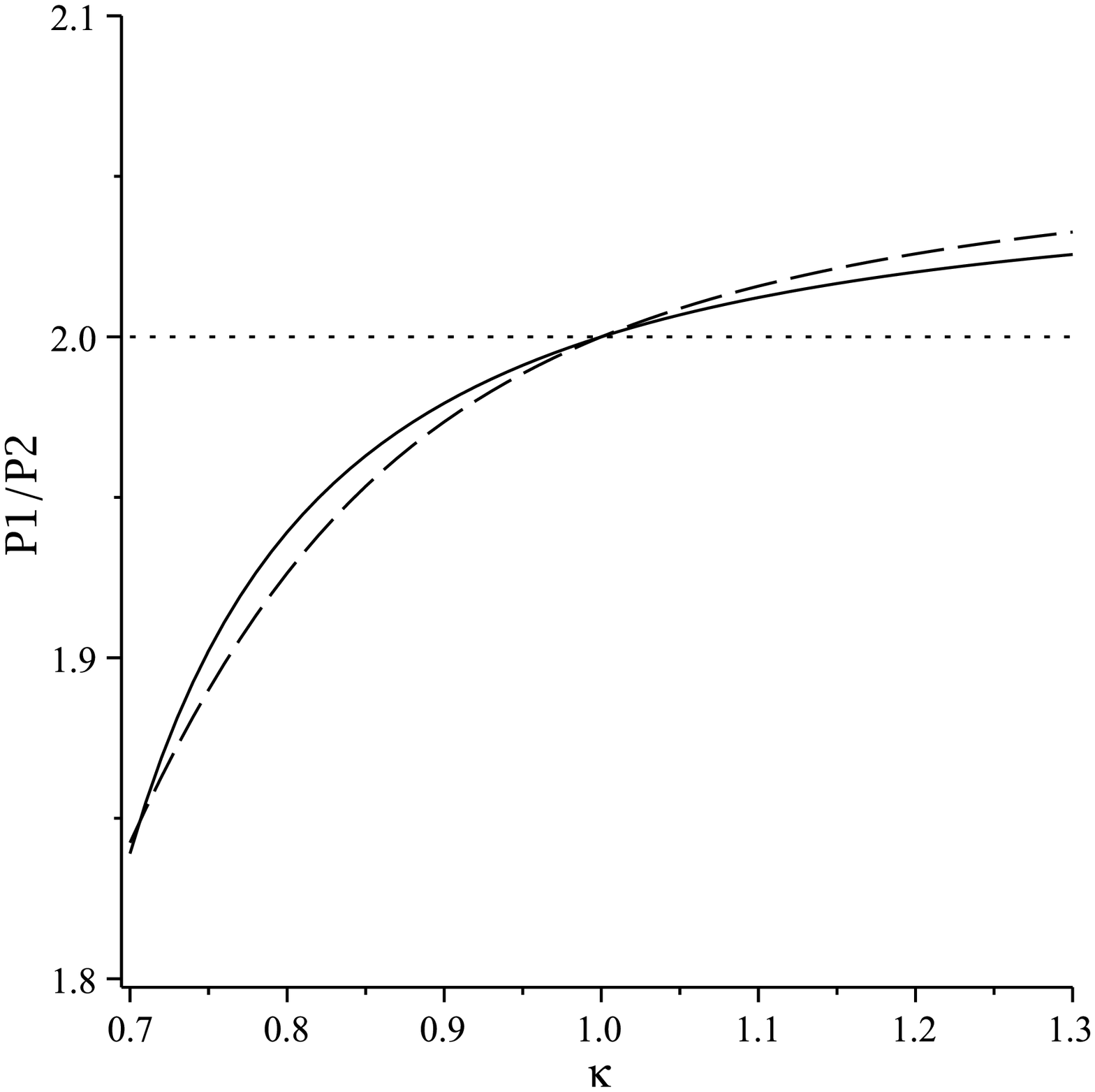}}

\caption{Equivalent to Figure 2 for Equation (25).}
\end{figure}

\section{Bead Propagation along Loop}
     \label{S-Bead Propagation along Loop}
Now, it is possible to add a time-dependence of the background state, representing a gas plug flowing along the magnetic field lines (or a bead moving along a string). It is assumed that the initial density profile of the bead is close to the end of the loop as in Section 4, before advancing along the loop. This can be written concisely as
\begin{equation}
\rho(z,t)=\left\{\begin{array}{ll}
\rho_{\mathrm{e}} & \quad{z\in[0,z_{\mathrm{0}}+v_{\mathrm{0}}(z)t-a]\cup[z_{\mathrm{0}}+v_{\mathrm{0}}(z)t+a,L]};\\
\rho_{\mathrm{b}} & \quad{z\in[z_{\mathrm{0}}+v_{\mathrm{0}}(z)t-a,z_{\mathrm{0}}+v_{\mathrm{0}}(z)t+a]},
\end{array}\right.
\end{equation}
where $t$ is time and $v_{\mathrm{0}}(z)$ is the flow velocity of the bead. We take the initial point of the bead close to $z_{\mathrm{0}}=L$ (and so $\epsilon\ll1$) and consider a constant speed of advancement in the negative $z$ direction, that is $v_{\mathrm{0}}(z)\equiv v_{\mathrm{0}},\, v_{\mathrm{0}} < 0$.  We now use the WKB method (see \opencite{Bender78} for more information on this technique) to solve the governing equation [Equation (1)] for the dynamic density profile. Application of the WKB method requires that the characteristic time of density changes should be small compared to the period of oscillations. This implies
\begin{equation}
\delta \equiv \frac{v_{\mathrm{0}}}{L}
\end{equation}
is small or $|\delta t| \ll 1$. This assumption is reasonable when compared to observations of flows (\opencite{Ruderman10} suggested a reasonable limit to flow velocities should be approximately $100$ km s$^{-1}$).

 Continuity of the solution at the endpoints of the bead, as in Sections 3 and 4, allows us to obtain the dispersion relationship
$$
2\gamma\alpha\tan(\gamma)\left(\tan\left(\frac{\omega{z_{\mathrm{0}}}}{v_{\mathrm{k,e}}}\right)+\Omega\right)\left(1-\tan\left(\frac{\omega{z_{\mathrm{0}}}}{v_{\mathrm{k,e}}}\right)\Omega\right)\left(1-\frac{1}{\kappa^2}\right)
$$
$$
+(\tan(\gamma)-2\gamma\alpha)\left[\left(\tan\left(\frac{\omega{z_{\mathrm{0}}}}{v_{\mathrm{k,e}}}\right)+\Omega\right)^2+\left(1-\tan\left(\frac{\omega{z_{\mathrm{0}}}}{v_{\mathrm{k,e}}}\right)\Omega\right)^2\right]
$$
\begin{equation}
+\frac{2\gamma\alpha}{\kappa^2}\left[\left(\tan\left(\frac{\omega{z_{\mathrm{0}}}}{v_{\mathrm{k,e}}}\right)+\Omega\right)^2+\kappa^2\left(1-\tan\left(\frac{\omega{z_{\mathrm{0}}}}{v_{\mathrm{k,e}}}\right)\Omega\right)^2\right]=0,
\end{equation}
where
\begin{equation}
\Omega=\tan\left(\frac{\omega{v_{\mathrm{0}}}t}{v_{\mathrm{k,e}}}\right).
\end{equation}

It is interesting to note that Equation (28) is equivalent to Equation (14) of \inlinecite{Soler11} who, however, investigated the properties of the fundamental harmonic of the kink mode within a solar prominence. Several key differences arise from those performed by \inlinecite{Soler11}, including, {\it e.g.}, the ratio of the density of the bead and loop, as explained in Section 2. Now, by making some further simplifications, we seek to solve Equation (28) analytically, hence allowing the explicit calculation of the $P1/P2$ ratio for the specific density profile studied in this article. 

Assuming that $L\gg|{v_{\mathrm{0}}t}|$ ({\it i.e.} at all modelled times, the distance travelled by bead is much less than $L$), in accordance with the WKB method, we find
\begin{equation}
\Omega\approx{\gamma\delta{t}}.
\end{equation}
Applying Equation (30) to Equation (28) and using the techniques exploited in Sections 3 and 4, we obtain that
\begin{equation}
\gamma_n=n\pi+\frac{2n^3\pi^3\alpha(\kappa^2-1)(\epsilon-\delta t)^2}{2n^2\pi^2\alpha(\kappa^2-1)[\delta t(6\epsilon-\delta t -1)+\epsilon(1-3\epsilon)]+\kappa^2f_1},
\end{equation}
where
\begin{equation}
f_1=1+n^2\pi^2\delta^2t^2+n^2\pi^2\epsilon^2.
\end{equation}

Here, we make note of several key limits. If bulk flows are absent, $v_{\mathrm{0}}=0$, we recover Equation (25) as we expect. For $\kappa<1$ and $t$ increasing, the ratio of the first two harmonics deviates from two towards zero, {\it i.e.}, as a bead which is heavier than the background loop moves through a thin tube, the ratio of the fundamental and first harmonics decreases. For $\kappa>1$ and $t$ increasing, we find that the period ratio {\it increases} from two, {\it i.e.}, a light bead travelling through a thin tube increases the value of the harmonic ratios. If $|v_{\mathrm{0}}t|\ll L$ we find that $P_1/P_2\approx{2},$ {\it i.e.}, a small bead travelling very slowly has little effect on the ratio of the periods of the first two harmonics, as we would expect.

Figure 4 shows how $\gamma_n$, calculated from Equation (31) (solid line) deviates from the canonical values $n\pi$ (dotted line) over time. We plot for $n=1,2,3$.  For larger values of $n$, the deviation from $n\pi$ occurs quicker, hence leading to a reduction in the $P1/P2$ ratio over time. Note, as $|v_{\mathrm{0}}t/L|\rightarrow{0.15}$, $\gamma_n$ slowly flattens out. This can be easily seen in Figure 4(b), where the higher harmonics show faster deviation from the idealised period ratio. It is interesting to note, that Figure 4(b) shows a non-linear deviation for the harmonics and appears, for $n=3$, to approach a minimum over time.

\begin{figure}[!tbp]
\includegraphics[scale=0.3]{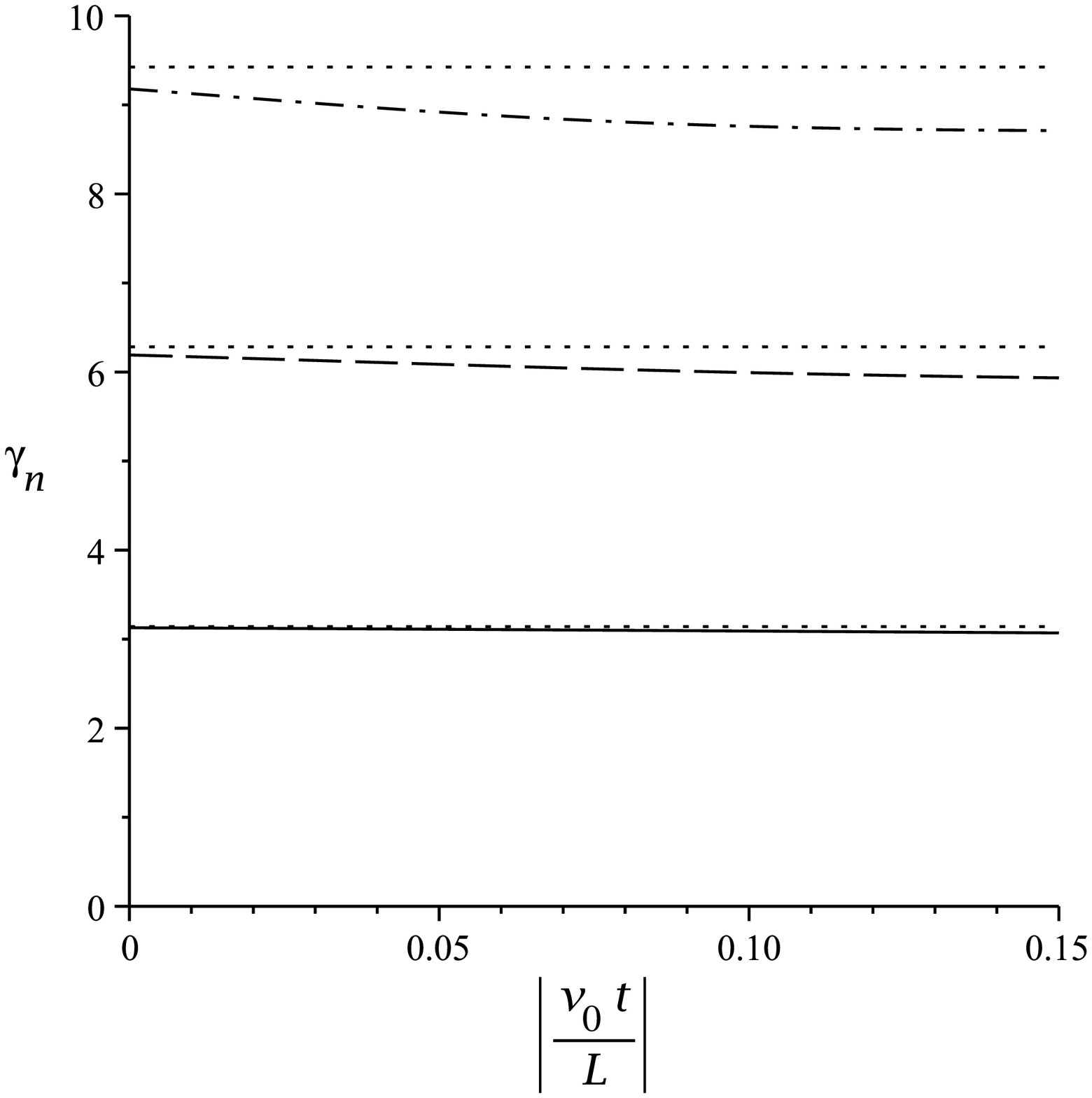}
\includegraphics[scale=0.3]{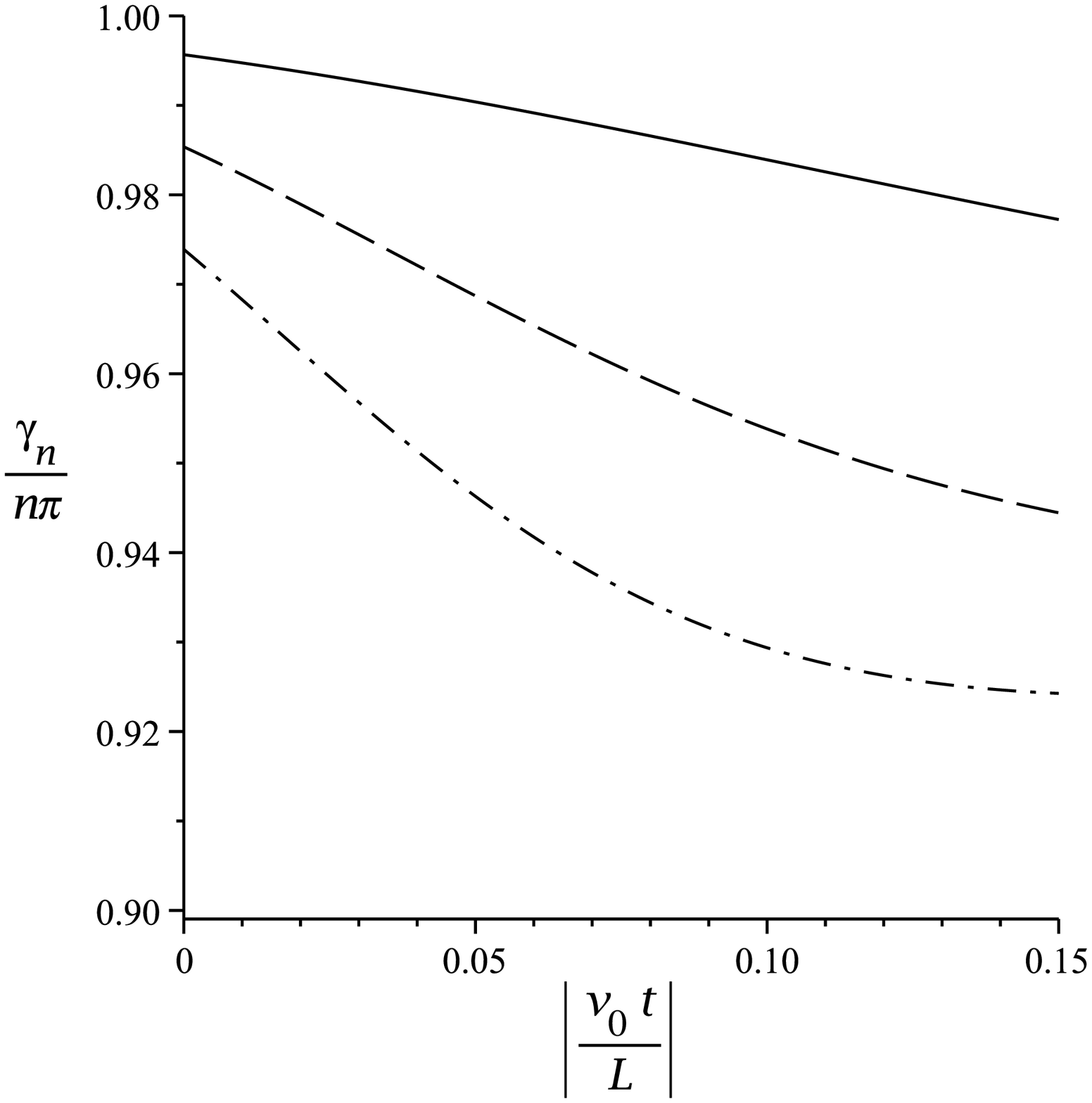}

\caption{(a) The deviation of $\gamma_n$ from $n\pi$ over time. (b) Ratio of $\gamma_n$ with respect to $n\pi$ over time. Line styles and corresponding harmonic are as follows: Solid line for $n$=1; dashed line for $n=2$ and dot-dashed line for $n=3$.}
\label{$Gamma_n$.}
\end{figure}

The influence of bulk flows from the footpoints of a loop into the corona, on standing MHD oscillations, is potentially an important tool for explaining the deviation of the observed $P1/P2$ ratios from the idealised, uniform oscillating loop ratios of two. In this article, we have derived an analytical approximation for the $P1/P2$ ratio of a moving bead (analogous to a small propagating density structure), finding approximate expressions which match closely with observations. It is interesting to note that, in the model presented here, the localised density enhancements do not need to propagate deeply into the corona to influence the period of standing oscillations. Strong bulk motions at the photospheric footpoints of such loop structures could cause rapid decrease of the $P1/P2$ ratio. In the following section, we discuss the influence of density stratification on the anti-nodes of standing MHD waves applicable to solar atmospheric oscillations, {\it e.g.}, to a coronal loop, thereby, presenting a second quantifiable value modified by a propagating density structure. Such study may open new avenues in spatio magneto-seismology, providing conditional constraints on coronal loop modelling.

\begin{figure}[!tbp]
\begin{center}
\rotatebox{0}{\includegraphics[scale=0.4]{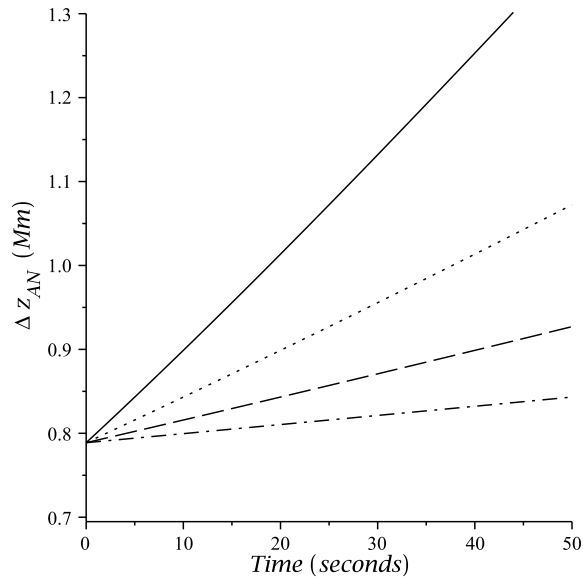}}
\end{center}

\caption{The change of the first harmonic anti-node shift over time with respect to velocities: $10$ km s$^{-1}$ (dot-dashed line); $25$ km s$^{-1}$ (dashed line); $50$ km s$^{-1}$ (dotted line) and $100$ km s$^{-1}$. Other parameters are $L=150$ Mm, $\epsilon=0.1$, $\kappa=0.9$, $n=2$ and $\alpha=0.05$.}
\label{Anti-node shifts.}
\end{figure}

\section{Anti-Node Shift}
    \label{S-Anti-Node Shift}
The shift of anti-nodes within loop oscillations has also been discussed in recent years (see, {\it e.g.}, \opencite{Verth07}; \opencite{Verth08a}). It has been found that the shift of the anti-nodes depends on both $\epsilon$ and $\kappa$ and that high density contrasts cause the largest deviation from canonical values, {\it i.e.}, from eigenfunctions of homogeneous loops. Here, discussion focuses on the influence that changing both the density distribution and the position of the discontinuity has on the anti-node shift.

The anti-node shift is an observable trait of oscillations which can be used as a novel diagnostic tool to infer background plasma properties. Using the anti-node shift to investigate solar structures is still in its infancy; mainly because the spatial resolution of current solar instrumentation is just about at the sensitivity to search for and detect this effect in standing oscillations. It is not yet thoroughly expanded upon in the literature. Work such as we present here could become a valuable addition to this new and exciting technique.

We shall begin by noting that the position of the anti-nodes can be found by calculating
\begin{equation}
\frac{{d}}{{d}z}(z_{\mathrm{AN}})=0,
\end{equation}
where $z=z_{\mathrm{AN}}$ is the positions of the anti-nodes.
To calculate shift, one must first calculate the position of the anti-nodes for a homogeneous loop. We then assume weak stratification $\rho_{\mathrm{e}} \approx \rho_{\mathrm{b}}$ to find the position of the anti-nodes for the inhomogenous case. Once these are known, it is simple to subtract one from the other to derive the anti-node shift expressed as
\begin{equation}
\Delta{z}_{\mathrm{AN}}=\left|\frac{(2n-1)L}{2}\left(\frac{\pi}{\gamma_n}-\frac{1}{n}\right)\right|.
\end{equation} 
Substituting Equation (20) into Equation (34), it is easy to calculate
\begin{equation}
\Delta z_{\mathrm{AN}}=\left|\frac{(2n-1)L}{2n} \left(\frac{3\epsilon^3n^2\pi^2(1-\kappa^2)+3\kappa^2(1+\epsilon^2n^2\pi^2)}{2\epsilon^3n^2\pi^2(1-\kappa^2)+3\kappa^2(1+\epsilon^2n^2\pi^2)}-1\right)\right|
\end{equation}
for the single-density-discontinuity case. Note that if $\kappa=1$, {\it i.e.}, there is no density jump, we retrieve $\Delta{z}_{\mathrm{AN}}=0$. Therefore, no change in the anti-nodes, as we would expect. Here, it is found that if there is a stronger contrast in the densities ({\it i.e.} $|\kappa|$ increases) then a higher anti-node shift would be manifested. Substituting characteristic values applicable to coronal loops, {\it e.g.}, $L=150$ Mm, $\epsilon=0.1$, $\kappa=0.9$, and $n=2$ ({\it i.e.}, the first harmonic) into Equation (35), we find $\Delta{z_{\mathrm{AN}}}=0.248$ Mm which is a relatively small deviation of node movement over the length of the loop. This would be harder to measure with the currently available instrumental limit, though with improving technology it is anticipated that such measurements should be carried out in the near future.

\begin{figure}[!tbp]
\begin{center}
\rotatebox{0}{\includegraphics[scale=0.4]{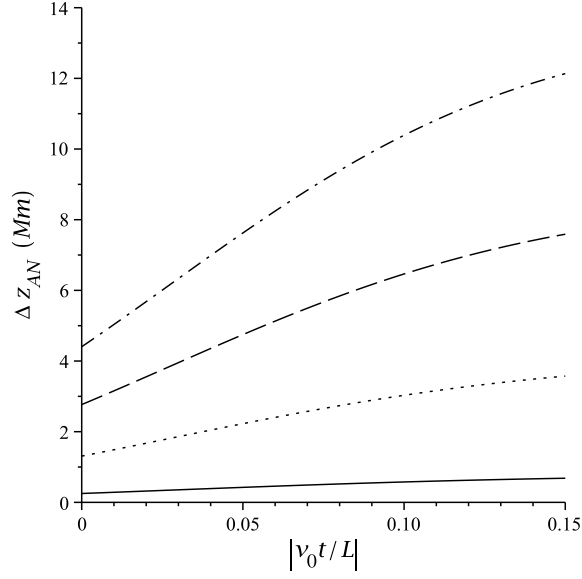}}
\end{center}

\caption{Variance of the anti-nodes over time for $n=2$. Different lines correspond to different sizes of $\alpha$ in our model, namely: $0.01$ (solid line), $0.05$ (dotted line), $0.1$ (dashed line), and $0.15$ (dot-dashed line). It is simple to see that larger beads cause larger shifts of the anti-nodes.}
\label{Anti-node shifts.}
\end{figure}

Next, let us derive the anti-node shift for the step-density. After some algebra we find
\begin{equation}
\Delta z_{\mathrm{AN}} =\left|\frac{(2n-1)L}{2n}\left(\frac{2n^2\pi^2\alpha\epsilon(\kappa^2-1)(3\epsilon-1)-\kappa^2(1+n^2\pi^2\epsilon^2)}{2n^2\pi^2\alpha\epsilon(\kappa^2-1)(2\epsilon-1)-\kappa^2(1+n^2\pi^2\epsilon^2)}-1\right)\right|.
\end{equation}
Again, for $\kappa={1}$ there is no anti-node shift as expected. Substituting $L=150$ Mm, $\epsilon=0.1$, $\kappa=0.9$,  , and $\alpha=0.05$ into Equation (36) returns $\Delta{z_{\mathrm{AN}}}=0.789$ Mm. This is a significant estimated deviation from the unperturbed anti-nodes, even close to the spatial resolution of the {\it Atmospheric Imaging Assembly} onboard the {\it Solar Dynamics Observatory} spacecraft (SDO/AIA) instrument, implying that observations of these shifts \underline{could} be possible. We would strongly encourage the community to pursue observational studies into spatio magneto-seismology.

Finally, we determine the anti-node shift of a thin loop over time when a bead is propagating away from its footpoint. This can be written as:
\begin{equation}
\Delta z_{\mathrm{AN}} =\left|\frac{(2n-1)L}{2n}\left(\frac{2n^2\pi^2\alpha(\kappa^2-1)f_2+\kappa^2f_1}{2n^2\pi^2\alpha(\kappa^2-1)\left[f_2+(\epsilon-\delta{t})^2\right]+\kappa^2f_1}-1\right)\right|
\end{equation}
where
\begin{equation}
f_2=\delta{t}(6\epsilon-\delta{t}-1)+\epsilon(1-3\epsilon)
\end{equation}
and $f_1$ is as described in Equation (32). 

In Figure 5, we plot the anti-node shift over time from $t=0$ (taken as the conditions stated after Equation (36) and varying $\alpha$, the ratio of half of the length of the bead to the length of the loop). The plot beautifully agrees with the intuitive result, that faster propagating bead will cause larger anti-node shift. Next, in Figure 6, anti-node shifts are plotted for different flowing bead lengths. It is found that small beads cause relatively small anti-node shifts (less than $2$ Mm). However, even for values such as $\alpha=0.15$, anti-node shifts of around $12$ Mm can be found. As the bead propagates away from the footpoint of the loop, it is found that the deviation from the unperturbed anti-nodes  increases. Given that the pixel size of the SDO/AIA instrument is fixed at $0.6$\arcsec, we note that values such as $2-12$ Mm may be easily observable with current techniques.

\section{Discussion and Conclusions}
     \label{S-Discussion and Conclusions}
In this article, we have presented both an analytical approximation for the period ratio of transversally oscillating coronal loops, with different density profiles representing a density jump, a static bead and a moving bead.  An estimate of the accompanying anti-node shift was found. These two quantifiable, and complimentary, physical parameters for transverse standing oscillations could provide clear insight into the background characteristics of magnetic loops in the solar atmosphere. The model discussed in this work encompasses localised bulk flows from loop footpoints into higher regions of the atmosphere. We find that these localised density enhancements can have a large and measurable influence by both modifying the $P1/P2$ period ratios and causing anti-node shifts within flux tube structures.

We began our analysis by studying the $P1/P2$ ratios of loops which had non-homogeneous density profiles. The dispersion relationship for each case was derived before analytical solutions were obtained. It was found, as suggested by, for example, \inlinecite{Diaz10} and \inlinecite{Soler11}, that a bead, heavier than the plasma in the background loop, will lead to a modification of the $P1/P2$ ratio to a value less than its canonical value. Although it is not directly studied in this article, it is interesting to note here that the Equations derived here also apply to beads {\it lighter} than the background density of the flux tube and show an increase of the $P1/P2$ period ratio from two. 

The model studied in this article was dependent on a number of parameters, specifically, the ratio of the kink speed of tube sections and $\epsilon$, the dimensionless position of the bead along the loop. Analytical solutions were obtained such that $\epsilon\ll{1}$, {\it i.e.}, a jet flow propagating from the photosphere or chromosphere into the corona (as observed by, {\it e.g.}, \opencite{Ofman08}). It is important to note that the background density of the loop studied in this model includes no stratification and, therefore, presents a simplistic example.

Wave propagation in a continuously stratifiied magnetic atmosphere, rather than a flux tube, has been studied extensively. Vertical and oblique magnetic fields were investigated by, {\it e.g.}, \inlinecite{FP58} and \inlinecite{LB79}, respectively. \inlinecite{Campos87} found that the P1/P2 ratio for standing Alfv\'{e}n waves in an isothermal atmosphere (where the layer is larger than the atmospheric scale height) is given by $j_2/j_1=2.25,$ where $j_{\mathrm{n}}$ is the $n$-th root of the Bessel function $J_{\mathrm{0}}$. This increase from 2 for Alfv\'{e}n waves is not in direct disagreement with our results for kink waves in a thin flux tube as each case neglects a different feature of a more complex, and realistic, model of a flux tube with continuous density stratification and flows. 

The $P1/P2$ period ratio was also derived as a function of time for a moving bead through a loop. It was found that larger values of $n$ cause quicker and larger divergence from the typical period of $n\pi$ leading to a decreasing period ratio initially. This suggests that the largest divergence from the expected $P1/P2$ period ratio for a homogeneous bounded flux tube occurs when the bead is in a spatial position such that $|v_{\mathrm{0}}t/L|\rightarrow{0.2}$. However, due to our approximation assumptions, we are unable to verify this analytically. We conclude that density stratification can play an important role in the deviation of the $P1/P2$ ratio from two. Future studies should be undertaken with the aim of discussing more advanced, and applicable, density profiles (such as a bead flowing through a non-homogeneous background loop).

Finally, anti-node shifts generated through density stratification were discussed. It was found that, as the bead propagates away from the end of the loop, the anti-node shifts increase in general. Overall, we suggest that localised bulk flows within loop structures could be an important factor in explaining observed localised anti-node shifts.

\acknowledgements
We first thank M.S. Ruderman for many excellent discussions and thoughts which improved this manuscript. We thank the Science and Technology Facilities Council (UK) and the School of Mathematics and Statistics, University of Sheffield (UK) for the support received. RE is thankful to the NSF, 
Hungary (OTKA, Ref. No. K83133) and acknowledges M. K\'eray for patient encouragement. Research at the Armagh Observatory is grant-aided by the N. Ireland Dept. of Culture, Arts and Leisure.

\bibliography{Influence_of_stratification_on_P1P2_ratios.bib}{}

\begin{thebibliography}{43}
\ifx \bisbn   \undefined \def \bisbn  #1{ISBN #1}\fi
\ifx \binits  \undefined \def \binits#1{#1}\fi
\ifx \bauthor  \undefined \def \bauthor#1{#1}\fi
\ifx \batitle  \undefined \def \batitle#1{#1}\fi
\ifx \bjtitle  \undefined \def \bjtitle#1{\textit{#1}}\fi
\ifx \bvolume  \undefined \def \bvolume#1{\textbf{#1}}\fi
\ifx \byear  \undefined \def \byear#1{#1}\fi
\ifx \bissue  \undefined \def \bissue#1{#1}\fi
\ifx \bfpage  \undefined \def \bfpage#1{#1}\fi
\ifx \blpage  \undefined \def \blpage #1{#1}\fi
\ifx \burl  \undefined \def \burl#1{\textsf{#1}}\fi
\ifx \href  \undefined \def \href#1#2{\textsf{#2}}\fi
\ifx \doiurl  \undefined \def
  \doiurl#1{\href{http://dx.doi.org/#1}{\textsf{#1}}}\fi
\ifx \betal  \undefined \def \betal{\textit{et al.}}\fi
\ifx \binstitute  \undefined \def \binstitute#1{#1}\fi
\ifx \bctitle  \undefined \def \bctitle#1{#1}\fi
\ifx \beditor  \undefined \def \beditor#1{#1}\fi
\ifx \bpublisher  \undefined \def \bpublisher#1{#1}\fi
\ifx \bbtitle  \undefined \def \bbtitle#1{\textit{#1}}\fi
\ifx \bedition  \undefined \def \bedition#1{#1}\fi
\ifx \bseriesno  \undefined \def \bseriesno#1{\textbf{#1}}\fi
\ifx \blocation  \undefined \def \blocation#1{#1}\fi
\ifx \bsertitle  \undefined \def \bsertitle#1{\textit{#1}}\fi
\ifx \bsnm \undefined \def \bsnm#1{#1}\fi
\ifx \bsuffix \undefined \def \bsuffix#1{#1}\fi
\ifx \bparticle \undefined \def \bparticle#1{#1}\fi
\ifx \barticle \undefined \def \barticle#1{}\fi
\ifx \botherref \undefined \def \botherref#1{}\fi
\ifx \url \undefined \def \url#1{\textsf{#1}}\fi
\ifx \bchapter \undefined \def \bchapter#1{}\fi
\ifx \bbook \undefined \def \bbook#1{}\fi
\ifx \bcomment \undefined \def \bcomment#1{#1}\fi
\ifx \oauthor \undefined \def \oauthor#1{#1}\fi
\ifx \citeauthoryear \undefined \def \citeauthoryear#1{#1}\fi
\def \endbibitem {}
\ifx \bconflocation  \undefined \def \bconflocation#1{#1} \fi

\bibitem[\protect\citeauthoryear{{Andries}, {Arregui}, and
  {Goossens}}{2005}]{Andries05}
\begin{barticle}
\bauthor{\bsnm{{Andries}}, \binits{J.}},
\bauthor{\bsnm{{Arregui}}, \binits{I.}},
\bauthor{\bsnm{{Goossens}}, \binits{M.}}:
\byear{2005},
\batitle{{Determination of the coronal density stratification from the
  observation of harmonic coronal loop oscillations}}.
\bjtitle{\apjl}
\bvolume{624},
\bfpage{L57}.
\end{barticle}
\endbibitem

\bibitem[\protect\citeauthoryear{{Andries} \textit{et~al.}}{2009}]{Andries09}
\begin{barticle}
\bauthor{\bsnm{{Andries}}, \binits{J.}},
\bauthor{\bsnm{{van Doorsselaere}}, \binits{T.}},
\bauthor{\bsnm{{Roberts}}, \binits{B.}},
\bauthor{\bsnm{{Verth}}, \binits{G.}},
\bauthor{\bsnm{{Verwichte}}, \binits{E.}},
\bauthor{\bsnm{{Erd{\'e}lyi}}, \binits{R.}}:
\byear{2009},
\batitle{{Coronal seismology by means of kink oscillation overtones}}.
\bjtitle{\ssr}
\bvolume{149},
\bfpage{3}.
\end{barticle}
\endbibitem

\bibitem[\protect\citeauthoryear{{Aschwanden}
  \textit{et~al.}}{1999}]{Aschwanden99}
\begin{barticle}
\bauthor{\bsnm{{Aschwanden}}, \binits{M.J.}},
\bauthor{\bsnm{{Fletcher}}, \binits{L.}},
\bauthor{\bsnm{{Schrijver}}, \binits{C.J.}},
\bauthor{\bsnm{{Alexander}}, \binits{D.}}:
\byear{1999},
\batitle{{Coronal loop oscillations observed with the Transition Region and
  Coronal Explorer}}.
\bjtitle{Astrophys. J.}
\bvolume{520},
\bfpage{880}.
\end{barticle}
\endbibitem

\bibitem[\protect\citeauthoryear{{Bender} and {Orszag}}{1978}]{Bender78}
\begin{bbook}
\bauthor{\bsnm{{Bender}}, \binits{C.M.}},
\bauthor{\bsnm{{Orszag}}, \binits{S.A.}}:
\byear{1978},
\bbtitle{{Advanced Mathematical Methods for Scientists and Engineers}},
\bfpage{484}.
\end{bbook}
\endbibitem

\bibitem[\protect\citeauthoryear{{Berghmans} and {Clette}}{1999}]{Berghmans99}
\begin{barticle}
\bauthor{\bsnm{{Berghmans}}, \binits{D.}},
\bauthor{\bsnm{{Clette}}, \binits{F.}}:
\byear{1999},
\batitle{{Active region EUV transient brightenings - First results by EIT of
  SOHO JOP80}}.
\bjtitle{\solphys}
\bvolume{186},
\bfpage{207}.
\end{barticle}
\endbibitem

\bibitem[\protect\citeauthoryear{{Campos}}{1986}]{Campos86}
\begin{bchapter}
\bauthor{\bsnm{{Campos}}, \binits{L.M.B.C.}}:
\byear{1986},
\bctitle{{On umbral oscillations as a sunspot diagnostic}}.
In: \bbtitle{Gough, D.O. (ed.)},
\bfpage{293}.
\end{bchapter}
\endbibitem

\bibitem[\protect\citeauthoryear{{Campos}}{1987}]{Campos87}
\begin{barticle}
\bauthor{\bsnm{{Campos}}, \binits{L.M.B.C.}}:
\byear{1987},
\batitle{{On waves in gases. Part II: Interaction of sound with magnetic and
  internal modes}}.
\bjtitle{Reviews of Modern Physics}
\bvolume{59},
\bfpage{363}.
\end{barticle}
\endbibitem

\bibitem[\protect\citeauthoryear{{De Moortel}}{2009}]{Demoortel09}
\begin{barticle}
\bauthor{\bsnm{{De Moortel}}, \binits{I.}}:
\byear{2009},
\batitle{{Longitudinal waves in coronal loops}}.
\bjtitle{\ssr}
\bvolume{149},
\bfpage{65}.
\end{barticle}
\endbibitem

\bibitem[\protect\citeauthoryear{{De Moortel} and {Brady}}{2007}]{DM07}
\begin{barticle}
\bauthor{\bsnm{{De Moortel}}, \binits{I.}},
\bauthor{\bsnm{{Brady}}, \binits{C.S.}}:
\byear{2007},
\batitle{{Observation of higher harmonic coronal loop oscillations}}.
\bjtitle{Astrophys. J.}
\bvolume{664},
\bfpage{1210}.
\end{barticle}
\endbibitem

\bibitem[\protect\citeauthoryear{{Deforest} and {Gurman}}{1998}]{Deforest98}
\begin{barticle}
\bauthor{\bsnm{{Deforest}}, \binits{C.E.}},
\bauthor{\bsnm{{Gurman}}, \binits{J.B.}}:
\byear{1998},
\batitle{{Observation of quasi-periodic compressive waves in solar polar
  plumes}}.
\bjtitle{\apjl}
\bvolume{501},
\bfpage{L217}.
\end{barticle}
\endbibitem

\bibitem[\protect\citeauthoryear{{D{\'{\i}}az}, {Oliver}, and
  {Ballester}}{2010}]{Diaz10}
\begin{barticle}
\bauthor{\bsnm{{D{\'{\i}}az}}, \binits{A.J.}},
\bauthor{\bsnm{{Oliver}}, \binits{R.}},
\bauthor{\bsnm{{Ballester}}, \binits{J.L.}}:
\byear{2010},
\batitle{{Prominence thread seismology using the $P_1/2P_2$ ratio}}.
\bjtitle{\apj}
\bvolume{725},
\bfpage{1742}.
\end{barticle}
\endbibitem

\bibitem[\protect\citeauthoryear{{Dymova} and {Ruderman}}{2005}]{Dymova05}
\begin{barticle}
\bauthor{\bsnm{{Dymova}}, \binits{M.V.}},
\bauthor{\bsnm{{Ruderman}}, \binits{M.S.}}:
\byear{2005},
\batitle{{Non-axisymmetric oscillations of thin prominence fibrils}}.
\bjtitle{\solphys}
\bvolume{229},
\bfpage{79}.
\end{barticle}
\endbibitem

\bibitem[\protect\citeauthoryear{{Edwin} and {Roberts}}{1983}]{Edwin83}
\begin{barticle}
\bauthor{\bsnm{{Edwin}}, \binits{P.M.}},
\bauthor{\bsnm{{Roberts}}, \binits{B.}}:
\byear{1983},
\batitle{{Wave propagation in a magnetic cylinder}}.
\bjtitle{\solphys}
\bvolume{88},
\bfpage{179}.
\end{barticle}
\endbibitem

\bibitem[\protect\citeauthoryear{{Erd{\'e}lyi}}{2006}]{Erdelyi06}
\begin{bchapter}
\bauthor{\bsnm{{Erd{\'e}lyi}}, \binits{R.}}:
\byear{2006},
\bctitle{{Proceedings of SOHO 18/GONG 2006/HELAS I, Beyond the spherical Sun}}.
In: \bbtitle{In: Fletcher, K., Thompson, M. (eds.)}
\bseriesno{624},
\bfpage{{15.1 (on CDROM)}}.
\end{bchapter}
\endbibitem

\bibitem[\protect\citeauthoryear{{Erd{\'e}lyi} and {Taroyan}}{2008}]{Erdelyi08}
\begin{barticle}
\bauthor{\bsnm{{Erd{\'e}lyi}}, \binits{R.}},
\bauthor{\bsnm{{Taroyan}}, \binits{Y.}}:
\byear{2008},
\batitle{{Hinode EUV spectroscopic observations of coronal oscillations}}.
\bjtitle{\aap}
\bvolume{489},
\bfpage{L49}.
\end{barticle}
\endbibitem

\bibitem[\protect\citeauthoryear{{Erd{\'e}lyi} and {Verth}}{2007}]{Erdelyi07}
\begin{barticle}
\bauthor{\bsnm{{Erd{\'e}lyi}}, \binits{R.}},
\bauthor{\bsnm{{Verth}}, \binits{G.}}:
\byear{2007},
\batitle{{The effect of density stratification on the amplitude profile of
  transversal coronal loop oscillations}}.
\bjtitle{\aap}
\bvolume{462},
\bfpage{743}.
\end{barticle}
\endbibitem

\bibitem[\protect\citeauthoryear{{Ferraro} and {Plumpton}}{1958}]{FP58}
\begin{barticle}
\bauthor{\bsnm{{Ferraro}}, \binits{C.A.}},
\bauthor{\bsnm{{Plumpton}}, \binits{C.}}:
\byear{1958},
\batitle{{Hydromagnetic waves in a horizontally Stratified atmosphere. V.}}
\bjtitle{\apj}
\bvolume{127},
\bfpage{459}.
\end{barticle}
\endbibitem

\bibitem[\protect\citeauthoryear{{Jess} \textit{et~al.}}{2008}]{Jess08}
\begin{barticle}
\bauthor{\bsnm{{Jess}}, \binits{D.B.}},
\bauthor{\bsnm{{Mathioudakis}}, \binits{M.}},
\bauthor{\bsnm{{Erd{\'e}lyi}}, \binits{R.}},
\bauthor{\bsnm{{Verth}}, \binits{G.}},
\bauthor{\bsnm{{McAteer}}, \binits{R.T.J.}},
\bauthor{\bsnm{{Keenan}}, \binits{F.P.}}:
\byear{2008},
\batitle{{Discovery of spatial periodicities in a coronal loop using automated
  edge-tracking algorithms}}.
\bjtitle{\apj}
\bvolume{680},
\bfpage{1523}.
\end{barticle}
\endbibitem

\bibitem[\protect\citeauthoryear{{Jess} \textit{et~al.}}{2009}]{Jess09}
\begin{barticle}
\bauthor{\bsnm{{Jess}}, \binits{D.B.}},
\bauthor{\bsnm{{Mathioudakis}}, \binits{M.}},
\bauthor{\bsnm{{Erd{\'e}lyi}}, \binits{R.}},
\bauthor{\bsnm{{Crockett}}, \binits{P.J.}},
\bauthor{\bsnm{{Keenan}}, \binits{F.P.}},
\bauthor{\bsnm{{Christian}}, \binits{D.J.}}:
\byear{2009},
\batitle{{Alfv\'en waves in the lower solar atmosphere}}.
\bjtitle{Science}
\bvolume{323},
\bfpage{1582}.
\end{barticle}
\endbibitem

\bibitem[\protect\citeauthoryear{{Kopp} \textit{et~al.}}{1985}]{Kopp85}
\begin{barticle}
\bauthor{\bsnm{{Kopp}}, \binits{R.A.}},
\bauthor{\bsnm{{Poletto}}, \binits{G.}},
\bauthor{\bsnm{{Noci}}, \binits{G.}},
\bauthor{\bsnm{{Bruner}}, \binits{M.}}:
\byear{1985},
\batitle{{Analysis of loop flows observed on 27 March, 1980 by the UVSP
  instrument during the Solar Maximum Mission}}.
\bjtitle{\solphys}
\bvolume{98},
\bfpage{91}.
\end{barticle}
\endbibitem

\bibitem[\protect\citeauthoryear{{Leroy} and {Bel}}{1979}]{LB79}
\begin{barticle}
\bauthor{\bsnm{{Leroy}}, \binits{B.}},
\bauthor{\bsnm{{Bel}}, \binits{N.}}:
\byear{1979},
\batitle{{Propagation of waves in an atmosphere in the presence of a magnetic
  field. I - The method}}.
\bjtitle{\aap}
\bvolume{78},
\bfpage{129}.
\end{barticle}
\endbibitem

\bibitem[\protect\citeauthoryear{{Mathioudakis}, {Jess}, and
  {Erd{\'e}lyi}}{2012}]{Mathioudakis13}
\begin{botherref}
\oauthor{\bsnm{{Mathioudakis}}, \binits{M.}},
\oauthor{\bsnm{{Jess}}, \binits{D.B.}},
\oauthor{\bsnm{{Erd{\'e}lyi}}, \binits{R.}}:
2012,
{Alfv\'en waves in the solar atmosphere}.
\textit{\ssr},
94.
doi:\doiurl{{10.1007/s11214-012-9944-7}}.
\end{botherref}
\endbibitem

\bibitem[\protect\citeauthoryear{{Morton} and {Erd{\'e}lyi}}{2009}]{Morton09}
\begin{barticle}
\bauthor{\bsnm{{Morton}}, \binits{R.J.}},
\bauthor{\bsnm{{Erd{\'e}lyi}}, \binits{R.}}:
\byear{2009},
\batitle{{Transverse oscillations of a cooling coronal loop}}.
\bjtitle{\apj}
\bvolume{707},
\bfpage{750}.
\end{barticle}
\endbibitem

\bibitem[\protect\citeauthoryear{{Nakariakov} and {Ofman}}{2001}]{Nakariakov01}
\begin{barticle}
\bauthor{\bsnm{{Nakariakov}}, \binits{V.M.}},
\bauthor{\bsnm{{Ofman}}, \binits{L.}}:
\byear{2001},
\batitle{{Determination of the coronal magnetic field by coronal loop
  oscillations}}.
\bjtitle{\aap}
\bvolume{372},
\bfpage{L53}.
\end{barticle}
\endbibitem

\bibitem[\protect\citeauthoryear{{Ofman} and {Wang}}{2008}]{Ofman08}
\begin{barticle}
\bauthor{\bsnm{{Ofman}}, \binits{L.}},
\bauthor{\bsnm{{Wang}}, \binits{T.J.}}:
\byear{2008},
\batitle{{Hinode observations of transverse waves with flows in coronal
  loops}}.
\bjtitle{\aap}
\bvolume{482},
\bfpage{L9}.
\end{barticle}
\endbibitem

\bibitem[\protect\citeauthoryear{{O'Shea} \textit{et~al.}}{2007}]{Oshea07}
\begin{barticle}
\bauthor{\bsnm{{O'Shea}}, \binits{E.}},
\bauthor{\bsnm{{Srivastava}}, \binits{A.K.}},
\bauthor{\bsnm{{Doyle}}, \binits{J.G.}},
\bauthor{\bsnm{{Banerjee}}, \binits{D.}}:
\byear{2007},
\batitle{{Evidence for wave harmonics in cool loops}}.
\bjtitle{\aap}
\bvolume{473},
\bfpage{L13}.
\end{barticle}
\endbibitem

\bibitem[\protect\citeauthoryear{{Roberts}}{2000}]{Roberts00}
\begin{barticle}
\bauthor{\bsnm{{Roberts}}, \binits{B.}}:
\byear{2000},
\batitle{{Waves and oscillations in the corona (invited review)}}.
\bjtitle{\solphys}
\bvolume{193},
\bfpage{139}.
\end{barticle}
\endbibitem

\bibitem[\protect\citeauthoryear{{Roberts}, {Edwin}, and
  {Benz}}{1984}]{Roberts84}
\begin{barticle}
\bauthor{\bsnm{{Roberts}}, \binits{B.}},
\bauthor{\bsnm{{Edwin}}, \binits{P.M.}},
\bauthor{\bsnm{{Benz}}, \binits{A.O.}}:
\byear{1984},
\batitle{{On coronal oscillations}}.
\bjtitle{Astrophys. J.}
\bvolume{279},
\bfpage{857}.
\end{barticle}
\endbibitem

\bibitem[\protect\citeauthoryear{{Rosenberg}}{1970}]{Rosenberg70}
\begin{barticle}
\bauthor{\bsnm{{Rosenberg}}, \binits{H.}}:
\byear{1970},
\batitle{{Evidence for MHD pulsations in the solar corona}}.
\bjtitle{\aap}
\bvolume{9},
\bfpage{159}.
\end{barticle}
\endbibitem

\bibitem[\protect\citeauthoryear{{Ruderman}}{2010}]{Ruderman10}
\begin{barticle}
\bauthor{\bsnm{{Ruderman}}, \binits{M.S.}}:
\byear{2010},
\batitle{{The effect of flows on transverse oscillations of coronal loops}}.
\bjtitle{\solphys}
\bvolume{267},
\bfpage{377}.
\end{barticle}
\endbibitem

\bibitem[\protect\citeauthoryear{{Ruderman} and
  {Erd{\'e}lyi}}{2009}]{Ruderman09}
\begin{barticle}
\bauthor{\bsnm{{Ruderman}}, \binits{M.S.}},
\bauthor{\bsnm{{Erd{\'e}lyi}}, \binits{R.}}:
\byear{2009},
\batitle{{Transverse oscillations of coronal loops}}.
\bjtitle{\ssr}
\bvolume{149},
\bfpage{199}.
\end{barticle}
\endbibitem

\bibitem[\protect\citeauthoryear{{Soler} and {Goossens}}{2011}]{Soler11}
\begin{barticle}
\bauthor{\bsnm{{Soler}}, \binits{R.}},
\bauthor{\bsnm{{Goossens}}, \binits{M.}}:
\byear{2011},
\batitle{{Kink oscillations of flowing threads in solar prominences}}.
\bjtitle{\aap}
\bvolume{531},
\bfpage{A167}.
\end{barticle}
\endbibitem

\bibitem[\protect\citeauthoryear{{Soler}, {Ruderman}, and
  {Goossens}}{2012}]{Soler12}
\begin{barticle}
\bauthor{\bsnm{{Soler}}, \binits{R.}},
\bauthor{\bsnm{{Ruderman}}, \binits{M.S.}},
\bauthor{\bsnm{{Goossens}}, \binits{M.}}:
\byear{2012},
\batitle{{Damped kink oscillations of flowing prominence threads}}.
\bjtitle{\aap}
\bvolume{546},
\bfpage{A82}.
\end{barticle}
\endbibitem

\bibitem[\protect\citeauthoryear{{Srivastava} \textit{et~al.}}{2008}]{Sri08}
\begin{barticle}
\bauthor{\bsnm{{Srivastava}}, \binits{A.K.}},
\bauthor{\bsnm{{Zaqarashvili}}, \binits{T.V.}},
\bauthor{\bsnm{{Uddin}}, \binits{W.}},
\bauthor{\bsnm{{Dwivedi}}, \binits{B.N.}},
\bauthor{\bsnm{{Kumar}}, \binits{P.}}:
\byear{2008},
\batitle{{Observation of multiple sausage oscillations in cool post-flare
  loop}}.
\bjtitle{\mnras}
\bvolume{388},
\bfpage{1899}.
\end{barticle}
\endbibitem

\bibitem[\protect\citeauthoryear{{Taroyan} and {Erd{\'e}lyi}}{2009}]{TR09}
\begin{barticle}
\bauthor{\bsnm{{Taroyan}}, \binits{Y.}},
\bauthor{\bsnm{{Erd{\'e}lyi}}, \binits{R.}}:
\byear{2009},
\batitle{{Heating diagnostics with MHD waves}}.
\bjtitle{\ssr}
\bvolume{149},
\bfpage{229}.
\end{barticle}
\endbibitem

\bibitem[\protect\citeauthoryear{{Uchida}}{1970}]{Uchida70}
\begin{barticle}
\bauthor{\bsnm{{Uchida}}, \binits{Y.}}:
\byear{1970},
\batitle{{Diagnosis of coronal magnetic structure by flare-associated
  hydromagnetic disturbances}}.
\bjtitle{\pasj}
\bvolume{22},
\bfpage{341}.
\end{barticle}
\endbibitem

\bibitem[\protect\citeauthoryear{{Van Doorsselaere}, {Nakariakov}, and
  {Verwichte}}{2007}]{VD07}
\begin{barticle}
\bauthor{\bsnm{{Van Doorsselaere}}, \binits{T.}},
\bauthor{\bsnm{{Nakariakov}}, \binits{V.M.}},
\bauthor{\bsnm{{Verwichte}}, \binits{E.}}:
\byear{2007},
\batitle{{Coronal loop seismology using multiple transverse loop oscillation
  harmonics}}.
\bjtitle{\aap}
\bvolume{473},
\bfpage{959}.
\end{barticle}
\endbibitem

\bibitem[\protect\citeauthoryear{{Verth} and {Erd{\'e}lyi}}{2008}]{Verth08a}
\begin{barticle}
\bauthor{\bsnm{{Verth}}, \binits{G.}},
\bauthor{\bsnm{{Erd{\'e}lyi}}, \binits{R.}}:
\byear{2008},
\batitle{{Effect of longitudinal magnetic and density inhomogeneity on
  transversal coronal loop oscillations}}.
\bjtitle{\aap}
\bvolume{486},
\bfpage{1015}.
\end{barticle}
\endbibitem

\bibitem[\protect\citeauthoryear{{Verth}, {Erd{\'e}lyi}, and
  {Jess}}{2008}]{Verth08b}
\begin{barticle}
\bauthor{\bsnm{{Verth}}, \binits{G.}},
\bauthor{\bsnm{{Erd{\'e}lyi}}, \binits{R.}},
\bauthor{\bsnm{{Jess}}, \binits{D.B.}}:
\byear{2008},
\batitle{{Refined magnetoseismological technique for the solar corona}}.
\bjtitle{\apjl}
\bvolume{687},
\bfpage{L45}.
\end{barticle}
\endbibitem

\bibitem[\protect\citeauthoryear{{Verth} \textit{et~al.}}{2007}]{Verth07}
\begin{barticle}
\bauthor{\bsnm{{Verth}}, \binits{G.}},
\bauthor{\bsnm{{Van Doorsselaere}}, \binits{T.}},
\bauthor{\bsnm{{Erd{\'e}lyi}}, \binits{R.}},
\bauthor{\bsnm{{Goossens}}, \binits{M.}}:
\byear{2007},
\batitle{{Spatial magneto-seismology: effect of density stratification on the
  first harmonic amplitude profile of transversal coronal loop oscillations}}.
\bjtitle{\aap}
\bvolume{475},
\bfpage{341}.
\end{barticle}
\endbibitem

\bibitem[\protect\citeauthoryear{{Verwichte}
  \textit{et~al.}}{2004}]{Verwichte04}
\begin{barticle}
\bauthor{\bsnm{{Verwichte}}, \binits{E.}},
\bauthor{\bsnm{{Nakariakov}}, \binits{V.M.}},
\bauthor{\bsnm{{Ofman}}, \binits{L.}},
\bauthor{\bsnm{{Deluca}}, \binits{E.E.}}:
\byear{2004},
\batitle{{Characteristics of transverse oscillations in a coronal loop
  arcade}}.
\bjtitle{\solphys}
\bvolume{223},
\bfpage{77}.
\end{barticle}
\endbibitem

\bibitem[\protect\citeauthoryear{{Wang}}{2011}]{Wang11}
\begin{barticle}
\bauthor{\bsnm{{Wang}}, \binits{T.}}:
\byear{2011},
\batitle{{Standing slow-mode waves in hot coronal loops: Observations,
  modeling, and coronal seismology}}.
\bjtitle{\ssr}
\bvolume{158},
\bfpage{397}.
\end{barticle}
\endbibitem

\bibitem[\protect\citeauthoryear{{Zhugzhda}}{1984}]{Zhugzhda84}
\begin{barticle}
\bauthor{\bsnm{{Zhugzhda}}, \binits{Y.D.}}:
\byear{1984},
\batitle{{Resonance oscillations in sunspots}}.
\bjtitle{Soviet Astronomy Letters}
\bvolume{10},
\bfpage{19}.
\end{barticle}
\endbibitem

\end{thebibliography}
\bibliographystyle{spr-mp-sola.bst}

\end{article}
\end{document}